# Investigating the migration of immiscible contaminant fluid flow in homogeneous and heterogeneous aquifers with high-precision numerical simulations

Alessandra Feo[1,2*], Fulvio Celico[1]

1 Department of Chemistry, Life Sciences and Environmental Sustainability, University of Parma, Parma, Italy

2 INFN Gruppo collegato di Parma, Parma, Italy

*e-mail address: alessandra.feo@unipr.it

## Abstract

Numerical modeling of the migration of three-phase immiscible fluid flow in variably saturated zones is challenging due to the different behavior of the system between unsaturated and saturated zones. This behavior results in the use of different numerical methods for the numerical simulation of the fluid flow depending on whether it is in the unsaturated or saturated zones. This paper shows that using a high-resolution shock-capturing conservative method to resolve the nonlinear governing coupled partial differential equations of a three-phase immiscible fluid flow allows the numerical simulation of the system through both zones providing a unitary vision (and resolution) of the migration of an immiscible contaminant problem within a porous medium. In particular, using different initial scenarios (including impermeable "lenses" in heterogeneous aquifers), three-dimensional numerical simulation results are presented on the temporal evolution of the contaminant migration following the saturation profiles of the three-phases fluids flow in variably saturated zones. It is considered either light nonaqueous phase liquid with a density less than the water, or dense nonaqueous phase liquid, which has densities greater than the water initially released in unsaturated dry soil. Our study shows that the fate of the migration of immiscible contaminants in variably saturated zones can be accurately described, using a unique mathematical conservative model, with different evolution depending on the value of the system's physical parameters, including the contaminant density, and accurately tracking the evolution of the sharp (shock) contaminant front.

# 1. Introduction

Multiphase flow problems refer to the simultaneous flow of two or more fluids separated by sharp interphases. They are observed in natural phenomena like multiphase flow within a porous structure, blood flow, nuclear reactions, oil and gas industry applications, etc. For these reasons, suitable numerical models are necessary to predict their physical behavior accurately. However, modeling multiphase fluid flow is challenging since a universally robust and accurate solution methodology has not yet been identified. It is important to mention that this work is only concerned with unsteady three-phase fluids flow in the variably saturated zone.

A multiphase fluid flow dynamics are governed by coupled conserved partial differential equations for each fluid flow, based on the Darcy Law and the mass and momentum conservation principles. They are written as a function of saturation, capillary pressure, permeability, porosity, density, and viscosity of each fluid flow. Since the capillary pressure and permeability of each phase is a function of the saturation, these equations are highly nonlinear and, due to the gravity and pressure gradients, are responsible for creating sharp (shocks) front and rarefaction, which may introduce significant errors in the numerical simulations. For this reason, it is not easy to obtain a numerical solution that converges to the physical solution, although there is an increasing effort to construct conservative numerical solution methods. See, for example, recent reviews for the numerical resolution of the Richards' equation for variably saturated flow [1-2], immiscible fluid flow in unsaturated zones [3], transport modeling in the heterogeneous porous medium [4-5], including nonaqueous phase liquid (NAPL) [6-7], and variably saturated zones [8-10]. Several two-phase flow models exist in the literature that uses high-resolution numerical schemes, such as high-resolution central upwind scheme [11]. They have been used as a numerical approximation to solve the two-phase fluid flow model [12]. Or the numerical solution of hyperbolic conservation laws such as a space-time Conservation Element-Solution Element (CESE) high-resolution scheme for computing the transport of a diffusing pollutant in shallow flows [13], using a Kinetic Flux Vector Splitting scheme [14], using a fifth-order Weighted Essentially Non-Oscillatory (WENO) scheme [15]. Finite volume WENO scheme and discontinuous Galerkin method have been used to solve multiphase flow models [16-20], a second-order accurate difference method for nonlinear conservation laws in three-dimensions [21], a shock-capturing numerical scheme for compressible multicomponent problems [22] and multi-medium flows [23]. Other examples of two-phase fluid flow applications for oil-water fluids flow can be found in [24-25].

The method used in this paper has been introduced in Ref. [26]. It is based on the high-resolution shock-capturing flux (HRSC) conservative method [11,27,28] to follow sharp discontinuities accurately and temporal dynamics of three-phase immiscible fluid flow in a porous medium. Several validation tests were performed in [26] to verify the accuracy of the HRSC method and the CactusHydro code. These tests include a comparison with an analytical model such as the Burgers' equation, the Buckley-Leverett model and a comparison with a two-dimensional unsaturated-saturated water flow model [10], together with the sand tank experimental data [29]. They show the absence of spurious oscillations in the solution and convergence to the "weak" solution as the grid is refined. The time evolution is performed using a forward in time explicit method rather than the most used, implicit one in which the discretization is based on a "backward in time" evolution. That requires the time step to be sufficiently small since the method is "conditionally stable". The implicit methods, in contrast, are "unconditionally stable" but very expensive from the computational point of view and may lead to mass balance errors. See for example [30-31] for two-phase flow using an implicit pressure and explicit saturation schemes, three-phase fluid flow [32], finite element methods [33-36].

This work shows numerical results on the fate of either light NAPL or dense NAPL initially released on a dry unsaturated zone that migrates, depending on its density to the saturated aquifer one. In the literature, these scenarios are treated very differently. Here different initial scenarios are investigated (including impermeable "lenses" in heterogeneous aquifers). It is shown that using a unique mathematical model that uses an HRSC method, it is possible to numerical simulate the system in both variable saturated zones, giving a unitary vision and numerical resolution of the migration of immiscible contaminants in a porous medium. In particular, it follows the temporal (and three-dimensional spatial) evolution of the saturation profile of the three-phases fluid along the variably saturated zone and their conservation law all over the simulation.

## 2. Methods

## 2.1 Governing equations of three-phase immiscible fluid flow

This section briefly describes the numerical model introduced in Ref. [26]. Consider a three-dimensional three-phase fluid flow in a porous medium composed of nonaqueous (n), water (w), air (a), and a variably saturated zone. Using the conservation equation for the mass and momentum for each fluid phase together with the Darcy's velocity for each phase, we obtain the governing coupled partial differential equations (PDEs) for each phase-fluid flow,

$$\frac{\partial}{\partial t}(\rho_n \phi S_n) = \frac{\partial}{\partial x^i}\left[\rho_n \frac{k_{rn}}{\mu_n}k^{ij}\left(\frac{\partial p_a}{\partial x^j} + \rho_n g \frac{\partial z}{\partial x^j}\right)\right] - \frac{\partial}{\partial x^i}\left[\rho_n \frac{k_{rn}}{\mu_n}k^{ij}\left(\frac{\partial p_{can}}{\partial x^j}\right)\right] + q_n, \qquad (1)$$

$$\frac{\partial}{\partial t}(\rho_w \phi S_w) = \frac{\partial}{\partial x^i}\left[\rho_w \frac{k_{rw}}{\mu_w}k^{ij}\left(\frac{\partial p_a}{\partial x^j} + \rho_w g \frac{\partial z}{\partial x^j}\right)\right] - \frac{\partial}{\partial x^i}\left[\rho_w \frac{k_{rw}}{\mu_w}k^{ij}\left(\frac{\partial p_{caw}}{\partial x^j}\right)\right] + q_w, \qquad (2)$$

$$\frac{\partial}{\partial t}(\rho_a \phi S_a) = \frac{\partial}{\partial x^i}\left[\rho_a \frac{k_{ra}}{\mu_a}k^{ij}\left(\frac{\partial p_a}{\partial x^j} + \rho_a g \frac{\partial z}{\partial x^j}\right)\right] + q_a, \qquad (3)$$

where $x^i = (x, y, z)$ are the spatial cartesian coordinates, and $t$ is the time coordinate, $\rho_\alpha$ is the density $\left[\frac{M}{L^3}\right]$ of each phase, $\alpha = (w, n, a)$, $\mu_\alpha$ is the dynamics viscosity $\left[\frac{M}{LT}\right]$ of phase $\alpha$, $p_\alpha$ is the phase pressure $\left[\frac{M}{T^2 L}\right]$, $k_{r\alpha}$ is the dimensionless relative permeability of phase $\alpha$, $k^{ij}$ is the absolute permeability tensor $[L^2]$, $g$ denotes the gravitational acceleration $\left[\frac{L}{T^2}\right]$, $z$ is the depth $[L]$, $q_\alpha$ is the mass source/sink $\left[\frac{M}{L}\right]$, $\phi$ is the porosity, $S_\alpha$ is the dimensionless volumetric saturation of phase $\alpha$ which satisfy the relation,

$$S_w + S_n + S_a = 1. \qquad (4)$$

In Eqs. (1-3) it is used the pressure $p_a$ (that is the air pressure when $S_a$ is different from zero), and it is considered the capillary pressure for the air-water phase, $p_{caw} = (p_a - p_w)$, and the capillary pressure for the air-nonaqueous phase, $p_{can} = (p_a - p_n)$, where it is substituted $p_w = p_a - p_{caw}$, and $p_n = p_a - p_{can}$. The third capillary pressure, the nonaqueous-water phase can be deduced from the other two, and is given by, $p_{cnw} = (p_n - p_w) = (p_{caw} - p_{can})$ (in contrast to Refs. [37,38], where the air gradient pressure is assumed negligible). The relative permeabilities $k_{rw}, k_{rn}$ and $k_{ra}$ and the capillary pressures are function of saturations, $k_{r\alpha} = k_{r\alpha}(S_a, S_n, S_w), p_{can} = p_{can}(S_a, S_n, S_w)$ and $p_{caw} = p_{caw}(S_a, S_n, S_w)$. This paper uses the van Genuchten model [39,40] for the air-water and the air-nonaqueous capillary pressure (see subsection 2.3). But it is worth noticing that different solutions choices correspond to different porous mediums. The numerical solution and the method applied here are not affected by any particular choice. The absolute permeability $k^{ij}$ depends on the properties of the porous medium. The porosity $\phi$ is a function of pressure and can be linearly approximated to, $\phi = \phi_0[1 + c_R(p - p_0)]$, where $c_R$ is the rock compressibility, $\phi_0$ is the porosity at $p_0$, which is considered the atmospheric, and $p$ is the pressure (that will be associated with $p_a$).

It is convenient to define the product of the porosity $\phi$ and the saturation for each phase as, $\sigma_w \equiv \phi S_w$, $\sigma_n \equiv \phi S_n$, $\sigma_a \equiv \phi S_a$. Then, Eq. (4) can be written as, $\sigma_a + \sigma_n + \sigma_w = \phi_0[1 + c_R(p - p_0)]$. The left-hand side of (1-3), assuming constant density-viscosity for each phase, becomes

$$\frac{\partial \sigma_a}{\partial t} + \frac{\partial \sigma_n}{\partial t} + \frac{\partial \sigma_w}{\partial t} = \phi_0 c_R \frac{\partial p}{\partial t} \, , \tag{8}$$

where

$$\frac{\partial \sigma_{(\alpha)}}{\partial t} = -\frac{\partial}{\partial x^i}\left[F^i_{(\alpha)}(S_w, S_n, S_a, p)\right] + \frac{\partial}{\partial x^i}\left[Q^i_{(\alpha)}(S_w, S_n, S_a, p)\right] \tag{9}$$

and

$$F^i_{(\alpha)}(S_w, S_n, S_a, p) = -\frac{k_{r(\alpha)}(S_w, S_n, S_a)}{\mu_{(\alpha)}} k^{ij}\left(\frac{\partial p}{\partial x^j} + \rho_\alpha g \frac{\partial z}{\partial x^j}\right) \tag{10}$$

do not depend on the spatial derivative of the saturation, while

$$Q^i_{(\alpha)}(S_w, S_n, S_a, p) = -\frac{k_{r(\alpha)}(S_w, S_n, S_a)}{\mu_{(\alpha)}} k^{ij} \frac{\partial p_{ca(\alpha)}(S_w, S_n, S_a)}{\partial x^j} \tag{11}$$

depends on the spatial derivative of the saturation. The PDEs system to be numerically resolved is composed by Eqs. (8,9) and the variables $p, \sigma_w, \sigma_n, \sigma_a$ together with the functional form of the relative permeabilities and capillary pressures, written in terms of the saturations (see subsection 2.3).

## 2.2 High-resolution shock-capturing conservative numerical method

The system of PDEs (8-9) governing a three-phase immiscible fluid is naturally formulated in terms of conservation laws. In this system, the dominant part (in the regime of interest) is the hyperbolic one (Eq. (10)) and is responsible for forming shocks and propagating the density fronts. Indeed, to accurately reproduce the dynamics of the density discontinuity in time, one needs to separate the methods used to discretize the propagation (hyperbolic) part from the diffusive (parabolic) part (Eq.(11)). This separation is of fundamental importance since explicit methods in time evolution should be preferred when the propagation dominates over the diffusion. In contrast, implicit (in time) methods should be preferred when diffusion dominates over propagation. The importance of using conservative formulations, i.e., methods based on conservation law, is due to two fundamental theorems. The first one is due to Lax and Wendorff [27] and the second one by Hou and LeFloch [28]. Namely, conservative numerical schemes, if convergent, do converge to the weak solution of the problem. The second theorem states that non-conservative numerical schemes do not converge to the correct solution if a shock wave (or discontinuity) is present in the flow. These two theorems state that if a conservative formulation is used, then it is guaranteed that the numerical solution will converge to the correct solution. On the contrary, if a conservative formulation is not used, the numerical solution will converge to the incorrect solution when the flow develops a discontinuity.

There is a vast literature on numerical methods that take advantage of the conservative nature of the equation and are referred to as HRSC methods that accurately reproduce the discontinuous features of the solutions. For the system studied in this paper, information on the characteristic is not explicitly available. Indeed, we are using a variant of the HRSC central schemed discussed in Kurganov and Tadmor (KT) [11] that also has the advantage of dealing with a diffusive part in the conservative equations that is usually assumed not present in other schemes. A comprehensive and almost complete discussion of most of the possible HRSC methods can be found in Ref. [41] dedicated, in a different context, to the study of Relativistic Hydrodynamics. Here is briefly described the details of the method in one spatial dimension and one component fluid. Its extension to more components and three dimensions is straightforward. The equation to be solved is,

$$\frac{\partial}{\partial t} u(x,t) = -\frac{\partial}{\partial x} \left[ F\big(u(x,t)\big) + Q\left( u(x,t), \frac{\partial u(x,t)}{\partial x} \right) \right] , \qquad (12)$$

where $F\big(u(x)\big)$ is the flux associated with the hyperbolic part, and $Q(u(x), du(x)/dx)$ is the parabolic part of the flux. Notice that it is not explicitly written the dependence on $t$ in all quantities. How this decomposition is achieved for the system in discussion is shown in (10-11). The variable is discretized assuming, at any time, a given value $(u_i)$ on each grid point $(x_i)$. For intermediate points is assumed a piecewise-linear reconstruction inside each cell, $u_i(x) = u_i + s_i(x - x_i)$, where $s_i$ is the "slope" of the linear reconstruction, and they are constructed using total-variation-diminishing (TVD) properties (needed for the theorem named above) and based on the min-mod slope limiter [Kolgan 72, val Leer 79]. Indeed, the variable is assumed to have a jump discontinuity at the cell border, i.e., different values on the right and the left at the points $x_{i+1/2}$ called $u_i^+$ and $u_{i+1}^-$. This implies that the computed values of the flux $F(u(x))$ and $Q(u(x), du(x)/dx) = Q(u(x), u'(x))$ will be different at the cell boundary but these values may be used to compute the effective flux at the cell boundary $H_{j+1/2}$ and $P_{j+1/2}$. Here $H_{j+1/2}$ is a function of $F(u^+_{j+1/2})$ and $F(u^-_{j+1/2})$ while $P_{j+1/2}$ is constructed in terms of the values of $Q(x, u(x), u'(x))$. Once a prescription to construct this flux is given, one has a conservative flux method for the solution, namely one has just to numerical solve in time the ordinary differential equations (Methods of Lines) associated to each grid point:

$$\frac{d}{dt}u_j = -\frac{H_{j+1/2} - H_{j-1/2}}{\Delta x} + \frac{P_{j+1/2} - P_{j-1/2}}{\Delta x}. \tag{13}$$

The KT method amount to the following choice for the numerical fluxes:

$$H_{j+1/2} = -\frac{1}{2}\left[F(u^+_{j+1/2}) + F(u^-_{j+1/2})\right] - \frac{1}{2}a_{j+1/2}\left[u^+_{j+1/2} - u^-_{j+1/2}\right], \tag{14}$$

$$P_{j+1/2} = \frac{1}{2}\left[Q\left(u_j, \frac{u_{j+1} - u_j}{\Delta x}\right) + Q\left(u_{j+1}, \frac{u_{j+1} - u_j}{\Delta x}\right)\right], \tag{15}$$

where the $a_{j+1/2}$ depends on the explicit functional dependence of the flux $F(x, u(x))$ that, in this case, is not explicitly known without assuming an explicit dependence on the permeabilities and their derivatives. However, independently on the choice of the permeabilities, the flux on the two sides always have the same sign and indeed, to use for the $H_{j+1/2}$ the values computed on the "left" or on the "right" depend on the sign of $F(x, u(x))$ at the interface between two adjacent cells, namely $H_{j+1/2} = -F(u^+_{j+1/2})$ if it is negative, while $H_{j+1/2} = F(u^-_{j+1/2})$ if it is positive.

The Kurganov-Tadmor (KT) scheme avoids the local Riemann problem. The method belongs to the Monotonic Upstream-centered Scheme for Conservation Law (MUSCL) class suggested by van Leer in 1973. The KT scheme archives second-order accuracy in space using analytical information on the flux form. However, the variation of the method used here that uses the first-order upwind formula for the fluxes and the min-mod flux limiter avoids using any analytical information (besides monotonicity) and, indeed, only the point values of the flux are required. This is of great advantage since it can use tabulated values for permeabilities. The only penalty is that it is a first-order accuracy at the discontinuities and over the whole simulation grid. This is not a big drawback since one must keep in mind that any method will be first-order accuracy at the physical discontinuities, and the method is targeted to study and follow the evolution of sharp discontinuity fronts.

This technique requires the time step size to be sufficiently small and thus the use of High-Performance Computing (HPC). In Ref. [26], we implemented the innovative method in CactusHydro, based on the Cactus computational toolkit [42-44], an open-source software framework for developing parallel HPC simulation codes where data are evolved on a cartesian mesh with possible refinement levels using Carpet [45-46]. See Refs. [47-49], PFLOTRAN, an open source, massively parallel subsurface flow and reactive transport code [50-52], Parflow Hydrologic model for surface and subsurface flow on HPC [53-55], RichardsFOAM [56-57], CATHY (CATchment HYdrology [58], for related examples in HPC.

## 2.3 Three Permeabilities and capillary pressures model

The relative permeabilities for three phases are given in Ref. [39] and listed below,

$$k_{rw} = S_{ew}^{1/2} \left[ 1 - \left( 1 - S_{ew}^{1/m} \right)^m \right]^2, \tag{16}$$

$$k_{ra} = (1 - S_{et})^{1/2} \left( 1 - S_{et}^{1/m} \right)^{2m}, \tag{17}$$

$$k_{rn} = (S_{et} - S_{ew})^{1/2} \left[ \left( 1 - S_{ew}^{1/m} \right)^m - \left( 1 - S_{et}^{1/m} \right)^m \right]^2, \tag{18}$$

where $S_{et}$ is the total effective liquid saturation defined in terms of the irreducible wetting phase saturation $S_{wir}$. This paper uses the van Genuchten model [38] where the effective saturation, $S_e$, is given by, $S_e = \left[ 1 + \alpha p_c^n \right]^{\left( 1 - \frac{1}{n} \right)}$, and $\alpha$ and $n$ are model parameters. It can be solved for $p_c$,

$$p_c = -p_{c0} \left( 1 - S_e^{1/m} \right)^{1-m}, \tag{19}$$

where $n = \frac{1}{1-m}$, and $p_{c0} = \alpha^{-1}$ is the capillary pressure at $S_e = 0$. Since $p_{caw} = p_{can} + p_{cnw}$, the capillary pressures are given by,

$$p_{can} = -p_{can0} \left( 1 - S_{et}^{1/m} \right)^{1-m} \tag{20}$$

$$p_{caw} = -p_{can0} \left( 1 - S_{et}^{1/m} \right)^{1-m} - p_{cnw0} \left( 1 - S_{ew}^{1/m} \right)^{1-m}. \tag{21}$$

# 3. Three-dimensional immiscible fluid flow numerical simulations results

This section presents the results of seven different three-dimensional numerical simulations "real" examples of an immiscible contaminant released to the environment in the unsaturated zone that migrates toward the saturated (aquifer) under the effect of the gravity force; exploring different scenarios in both homogeneous and heterogeneous aquifer systems. It is considered a three-phase fluid model composed of water, air, and nonaqueous phase liquid (light or dense NAPL) and investigates the temporal evolution of the migration of the contaminant following the saturation profiles of each phase along the variably saturated zone. We assumed constant density and constant viscosity for each phase. The effects of volatilization/biodegradation and dissolution are not considered.

## 3.1 Migration and distribution of a continuous leak of NAPL

### 3.1.1 LNAPL numerical results

We first consider a continuous leak source (with a spill rate of $1.15\ kg/s$) of immiscible LNAPL, placed on top of a parallelepiped at z = $[8.0, 10.0]$ m, $x = [-5.0, 5.0]m$, and $y = [-5.0, 5.0]m$, at $t = 0\ s$, as shown in Fig 1 (red box). The variably saturated zone grid geometry is assumed to be a parallelepiped of $80\ m$ long from $x = [-40, +40]$ (left-hand side), $32\ m$ wide from $y = [-16, +16]m$ (right-hand side), and $22\ m$ depth, $z = [+10, -12]m$, with a spatial resolution of $dx = dy = dz = 0.50\ m$, and a time step $dt = 0.025\ s$. The dimension of the grid has been chosen to cover the minimum possible part of the aquifer but big enough to avoid the boundaries effect (finite grid size) on the dynamics of the contaminant. The porous medium is composed of an unsaturated dry zone (air-NAPL) and a saturated one (filled up with water, colored in blue) separated by the groundwater table located at $z = 0.0\ m$. The numerical grid is oriented such that the gravity force is directed downwards and the y-axis is horizontal (with respect to the gravity). At the same time, the x-axis forms an angle of 15 degrees with respect to the horizon, and indeed, the gravity force forms an angle of 15 degrees with respect to the z-axis. Consequently, that originates a pressure gradient, and the flow goes toward the negative direction of the x-axis (see the blue arrows directed to the left-hand side).

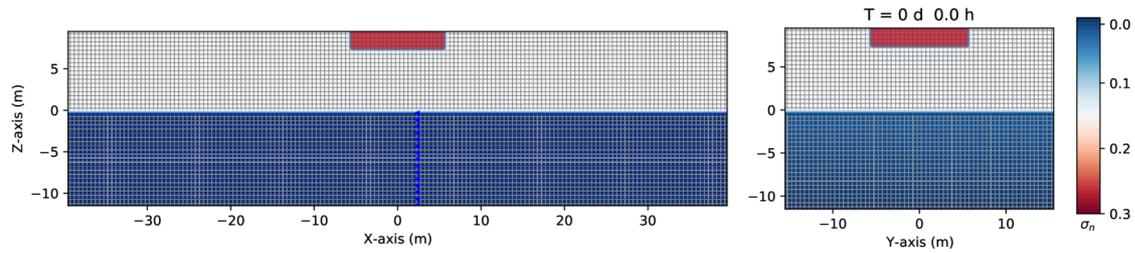

**Fig 1. Three-dimensional numerical grid geometry.** Example of the grid geometry used in the numerical simulation of a three-phase fluid flow (water + LNAPL + air) with a spatial grid resolution of $0.50\ m$ and a grid dimension of $80\ m \times 32 \times 22\ m$, at the initial time $t\ =\ 0\ s$. The red box is the continuous source of an immiscible contaminant at the top of the parallelepiped in the $z - x$ plane (left-hand side) and the $z - y$ plane (right-hand side), respectively.

All boundary conditions are no-flow except for the infiltration zone on top of the parallelepiped. A continuous source of contaminant indicates that the value of the saturation, $\sigma_n = S_n\ \phi$, remains constant on top of the grid along with the transient simulation. The legend at the right-hand side of Fig 1 indicates the saturation contour values of the contaminant in color bars.

Table 1 gives the material properties and parameter details used in the numerical simulations performed with CactusHydro [26] all over this work. In particular, it shows the density of the contaminant, which is $881 kg/m^3$ for the LNAPL, and $1200 kg/m^3$ for the DNAPL (see next subsection). It is important to stress that the numerical simulations with LNAPL and DNAPL differ only by their density values. The assigned input parameters used in the numerical simulations model have been taken from the Literature (not by conducting an experiment). They have been chosen to be representative of an aquifer in a "Loamy Sand" geological structure. The value of the density of $881 kg/m^3$ for the LNAPL corresponds to a value compatible with Crude Oil, while $1200 kg/m^3$ for the DNAPL was chosen in such a way to represent a generic, not too dense, one.

The porous medium is fixed to a value of porosity of 0.3. The absolute permeability is assumed to be $k = 4.14 \times 10^{-10} m^2$ all over the grid, with $k_x = k_y = k_z$. The relative permeabilities and capillary pressure were obtained using Eqs. (16-21). At zero saturation, the capillary pressure air-water is $793.80\ Pa$ or, equivalently, $0.081\ m$, while the capillary pressure air-nonaqueous at zero saturation is $556.66\ Pa$ or, equivalently, $0.0566\ m$.

| Parameter | Simbol | Value |
|---|---|---|
| Absolute permeability, $m^2$ | k | $4.14 \times 10^{-10}$ |
| Rock compressibility, $Pa^{-1}$ | $c_R$ | $4.35 \times 10^{-7}$ |
| Porosity | $\phi_0$ | 0.3 |
| Water viscosity, $kg/(ms)$ | $\mu_w$ | $10^{-3}$ |
| Water density, $kg/m^3$ | $\rho_w$ | $10^3$ |
| Oil viscosity, $kg/(ms)$ | $\mu_n$ | $10^{-1}$ |
| Oil density (LNAPL), $kg/m^3$ | $\rho_n$ | 881 |
| Oil density (DNAPL), $kg/m^3$ | $\rho_n$ | 1200 |
| Air viscosity, $kg/(ms)$ | $\mu_a$ | 0.000018 |
| Air density, $kg/m^3$ | $\rho_a$ | 1.225 |
| Van Genuchten parameters | $(n, m)$ | $(2, 1/2)$ |
| Irreducible wetting phase saturation | $S_{wir}$ | 0.057 |
| Capillary pressure air-water at zero saturation, $m$ | $p_{caw0}$ | 0.081 |
| Capillary pressure air-nonaqueous at zero saturation, $m$ | $p_{can0}$ | 0.0566 |
| Resolution, $m$ | $\Delta x = \Delta y = \Delta z$ | 0.5 |

**Table 1.** List of parameters used for the three-dimensional numerical simulations of LNAPL and DNAPL in variably saturated zones.

After being released into the environment, the LNAPL starts to migrate downward the unsaturated zone under the influence of gravity (see Fig 2, the left-hand side, where the saturation ($\sigma_n$) contours are shown for different times). It moves predominantly downward. Lateral spreading may also occur due to the effect of the capillary pressures. Fig 2 second row, the left-hand side, shows the impact of the contaminant after 5.7 hours. The contaminant arrives at the groundwater table after two days and 8.9 hours (third row, the left-hand side), where a capillary pressure between air-contaminant and contaminant-water is present. Since the contaminant has a density lighter than the water ($\rho_n = 881 kg/m^3$, see Table 1), it remains floating in the groundwater table zone representing a physical barrier. Part of the contaminant remains in the capillary fringe, part of it moves to the left direction due to a pressure gradient created by the gravity force's oblique position (see Fig 2 fourth row, left-hand side). Fig 2 also shows a few streamlines (in blue). Initially, they move undisturbed toward the left direction (left-hand side of Fig 2). They change their direction slightly under the contaminant's presence, which initially goes a bit down the groundwater table since the spill is sufficiently large, being a continuous leak source. On the right-hand side of Fig 2, the saturation contours are viewed in the $y - x$ plane. The transient numerical simulation shows the behavior of this LNAPL after 9 days and 11.6 hours, although it is possible to go further in the numerical simulation. The first two rows (on the right-hand side) show no contaminant since the plane is located at $z = 0\ m$ (groundwater table). Only when the contaminant reaches this level may it be noticed. See the third row on the right-hand side how the contaminant arrives at the groundwater table, keeping its initial square-like section. After a while, this becomes an oval-like shape (third and fourth row, right-hand side).

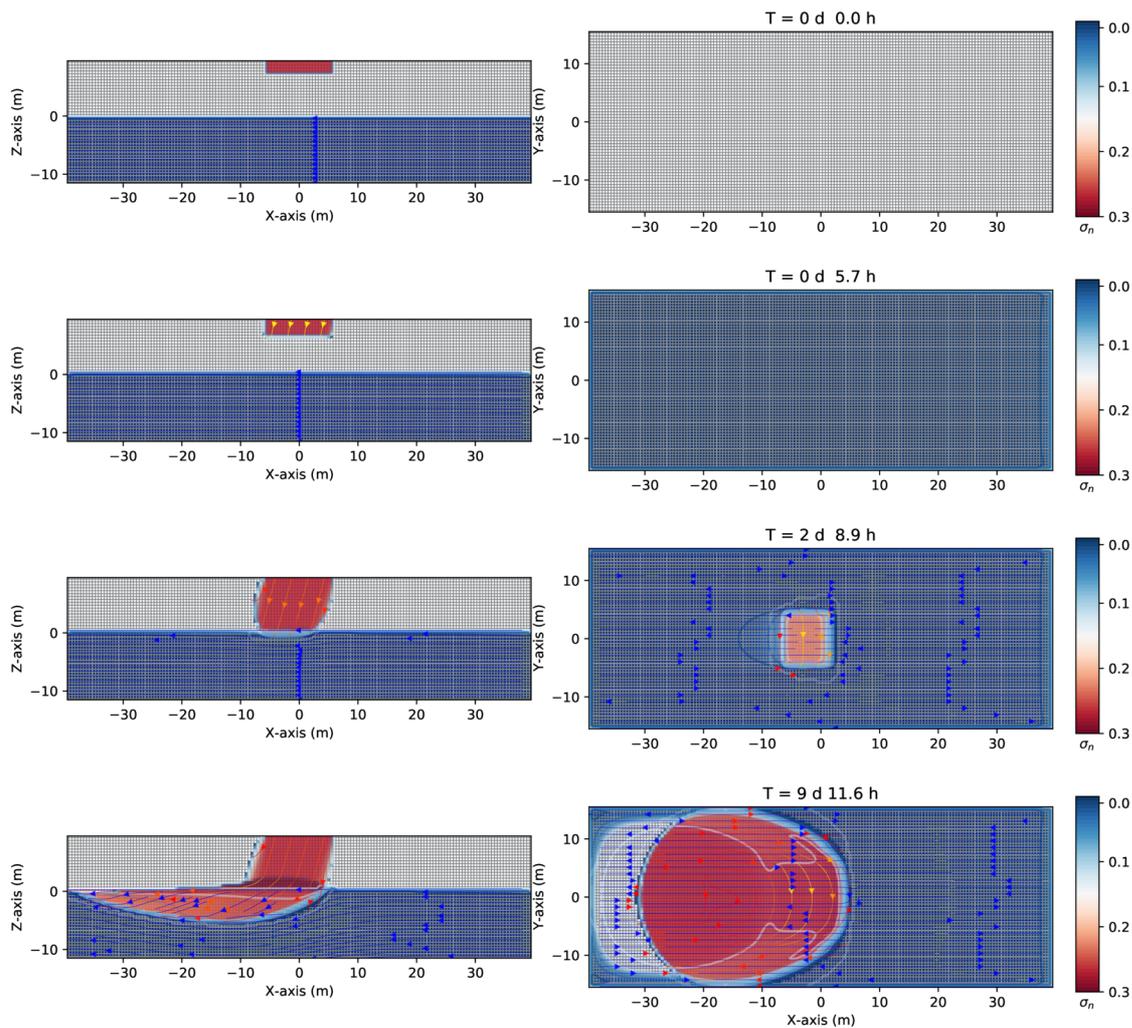

**Fig 2. Saturation contours of LNAPL (continuous source) at different times.** Three-dimensional numerical results on the saturation contours ($\sigma_n = S_n \phi$) of a three-phase immiscible fluid flow (water + a continuous source of LNAPL + air) using a spatial grid resolution of 0.50 $m$ and a grid dimension of 80 $m$ × 32 $m$ × 22 $m$, at different times. Left-hand side shows the saturation contours in the ($z - x$) plane. Right-hand side shows the saturation contours in the ($y - x$) one. Notice how the LNAPL moves in the unsaturated zone, although initially, it goes a bit down with respect to the groundwater table (-5 m) being a continuous contaminant spill. On the other hand, the first two rows show no contamination since the plane is located at the groundwater table.

Fig 3 shows three-dimensional numerical simulation results of the depth as a function of the water saturation $S_w$ (blue points), LNAPL saturation $S_n$ (red points), and air saturation $S_a$ (green points), at various times, for the continuous leak of LNAPL shown in Fig 2. Initially, at $t = 0\ s$ (left-hand side), there is a continuous source of LNAPL located at $z = 8.0\ m$, whose saturation is $S_n = 0.9$ (red points). Below $z = 8.0\ m$, this saturation goes abruptly to zero since initially the contaminant is situated on top of the parallelepiped (see also Fig 2, top, left-hand side). The water saturation (blue points), instead, is zero in the unsaturated zone (being completely dry) and is equal to one from $z = 0.0\ m$ to bellow (saturated zone). The air saturation (green points) equals $0.1$ in the unsaturated region where there is also the contaminant, increases to one between $z < 8.0\ m$ and the groundwater table, and goes to zero in the saturated zone (below $z = 0.0\ m$). Notice that the sum of the three phases' saturation is always one at any depth value.

After two days and $8.9$ hours, the LNAPL continuous leak source arrives at the groundwater table (see Fig 2 third row, left-hand side). See Fig 3 (middle), where the saturation of the LNAPL is kept constant to 0.9 (being a constant leak). It then arrives at the groundwater table (see red points), where it increases to 1.0 since the contaminant accumulates around $z = 0.0\ m$ before starting to move to the left-hand side together with the flow, to finally decrease very rapidly after passing through to the saturated zone where the sharp front goes to zero. Notice that the air saturation is now 0.1 above the groundwater table and zero in the saturated zone. After nine days and 11.6 hours (Fig 3 right-hand side, together with Fig 2 fourth row, left-hand side), the LNAPL remains partly in the capillary fringe with saturation one, partly goes into the saturated zone, being a continuous leak source (red points).

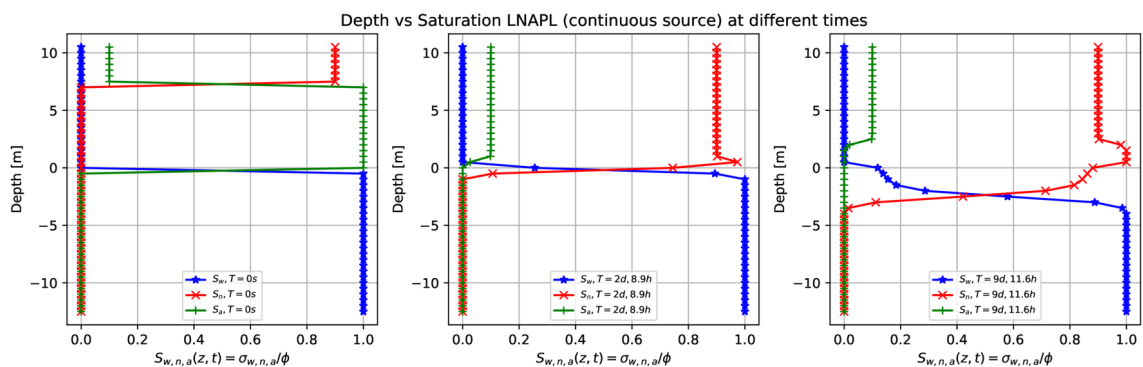

**Fig 3. Depth vs. saturation of LNAPL (continuous source) at different times.** Three-dimensional numerical simulation results of the depth as a function of the water saturation $S_w$ (blue points), LNAPL saturation $S_n$ (red points), and air saturation $S_a$ (green points) at various times for a continuous leak of LNAPL of Fig. 2. Initially, at t = 0 s, there is a sharp front of contaminant saturation situated on top of the grid, rapidly reaching zero when the height decreases. At the same time, it is filled by the air saturation (green point) in the unsaturated zone and water saturation in the saturated zone. Notice that the sum of the three-phase saturations is always one. For later times is being observed how the contaminant (red points) moves toward the saturated zone and remains floating while moving with the direction of the groundwater flow.

Fig 4 shows three-dimensional numerical results on the pressure for the LNAPL (continuous leak source) as a function of the depth at different times. Initially, at time equal to zero (blue points), the pressure is equal to the atmospheric one $(101325\,Pa)$, in the unsaturated zone composed by air-contaminant. Then the pressure increases, when moving from the groundwater table to the bottom, up to a value of $220\,KPa$. Similar behavior can be observed after two days and 8.9 hours and nine days and 11.6 hours (see the orange and green points, respectively). It can be noticed that the green points correspond to the contaminant that has already reached the groundwater table (see Fig 2). The pressure also increases slightly in the unsaturated zone.

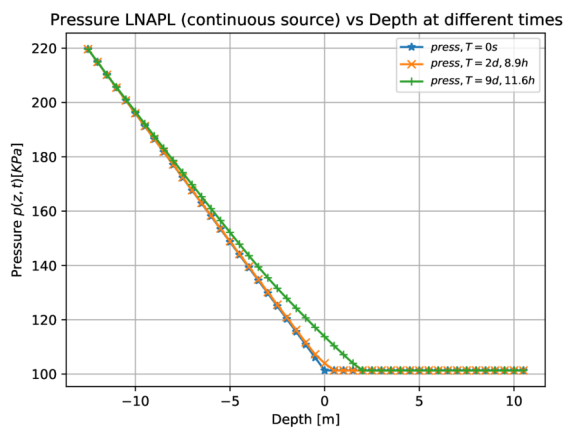

**Fig 4. Pressure vs. Depth for LNAPL (continuous source) at different times.** Three-dimensional numerical simulation results of the pressure as a function of the depth for a constant leak of LNAP at different times. Initially, at a time equal to zero (blue points), the pressure is equal to one atmosphere in the unsaturated zone composed by air-contaminant. It increases up to $220\,KPa$ in the saturated zone. For later times (green points), the pressure slightly increases also in the unsaturated zone when the LNAPL arrives at the saturated zone.

## 3.1.2 DNAPL numerical results

Fig. 5 shows similar numerical simulation results using the same set of values as the previous one (with LNAPL), except that the contaminant is a DNAPL with a density of $1200\,kg/m^3$ (see Table 1 for details). Fig 5 shows results of the saturation contours ($\sigma_n = S_n \phi$) of a three-phase immiscible fluid flow (water + a continuous source of DNAPL + air) using a spatial grid resolution of $0.50\,m$ and a grid dimension of $80\,m \times 32\,m \times 22\,m$, at different times. The left-hand side shows the saturation contours in the $(z-x)$ plane. The right-hand side shows the saturation contours in the $(z-y)$ one.

As in the previous case, the DNAPL is constantly released with a spill rate of $2.14\ kg/s$ in the unsaturated zone, being a continuous leak source. It goes downward due to gravity, first in the unsaturated dry zone (see Fig 5, first and second row, that show this movement at time zero, and after one day and 4.4 hours). Notice how the fluid flows in the left direction (Fig 5 left-hand side) due to a pressure gradient. Instead, the right-hand side shows the $z - y$ plane with a zero-gravity component effect and, thus, no privileged direction in the $y - $ axis. That is the reason why the right-hand side of Fig 5 is symmetric around the $y - $ axis. When the contaminant arrives at the groundwater table, which acts like a physical barrier created by the different phases, unlike the previous case with an LNAPL, it keeps moving to the bottom (aquiclude) of the saturated zone (due to its density) while moving in the left direction due to the pressure gradient. Notice how the DNAPL arrives at $-10\ m$ depth after four days and a few hours and advances more rapidly on the left-hand side to the previous LNAPL case (see Fig 5 the third and fourth row, left-hand side).

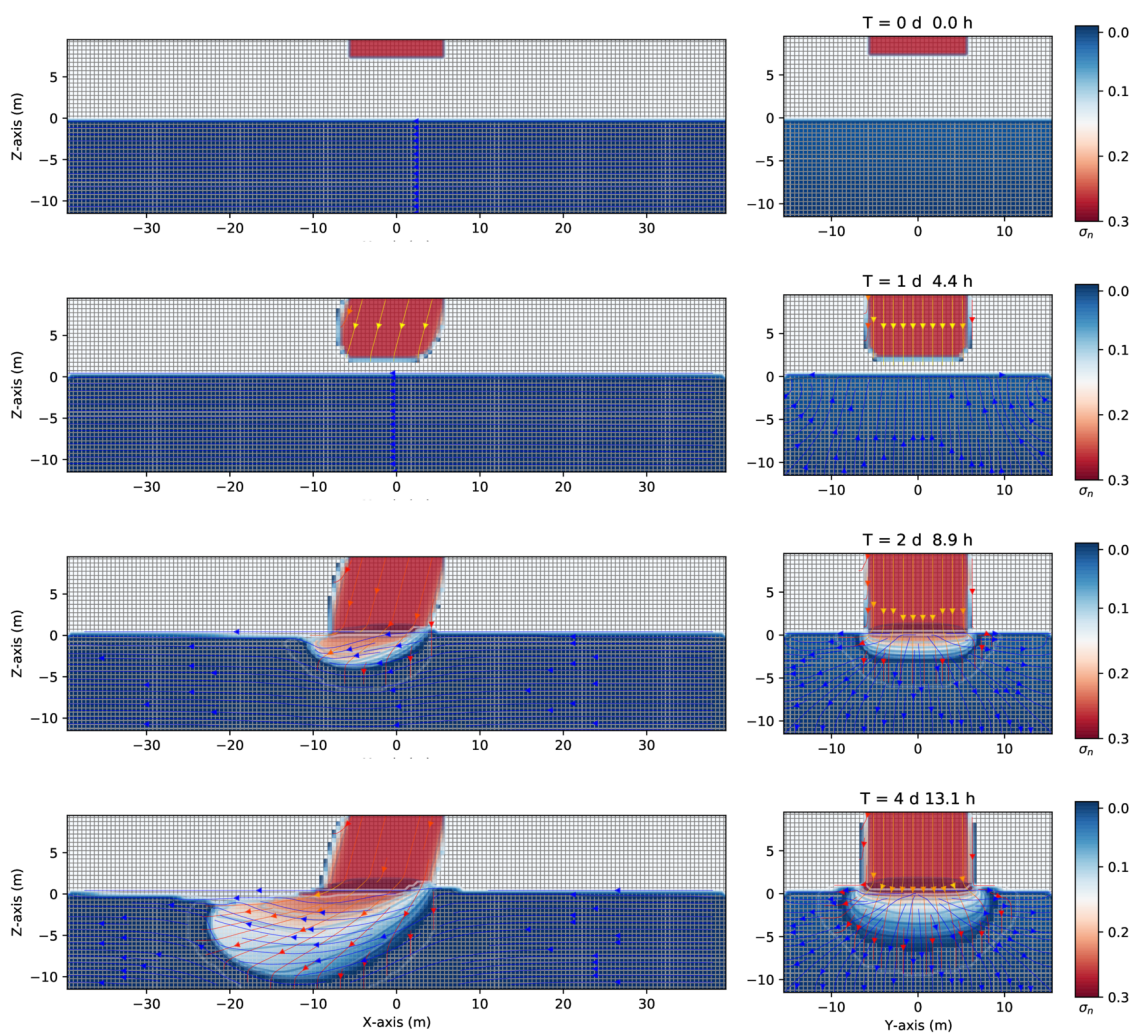

**Fig 5. Saturation contours of DNAPL (continuous source) at different times.** Three-dimensional numerical simulation results of the saturation contours ($\sigma_n = S_n \phi$) of a three-phase immiscible fluid flow (water + a continuous source of DNAPL + air) using a spatial grid resolution of 0.50 $m$ and a grid dimension of 80 $m$ × 32 $m$ × 22 $m$, at different times. The left-hand side shows the saturation contours in the $(z - x)$ plane. The right-hand side shows the saturation contours in the $(z - y)$ one. Notice how the DNAPL migrates through the saturated zone while moving in the left direction (where it is positioned a gravity at 15 degrees in the $z - x$ plane, left-hand side). The difference between the previous case (Fig 2) is that now the continuous leak of DNAPL keeps moving to the bottom (aquiclude) of the saturated zone while moving in the left direction due to the pressure gradient.

Fig 6 shows the three-dimensional numerical simulation results on the depth as a function of the water saturation $S_w$ (blue points), DNAPL saturation $S_n$ (red points), and air saturation $S_a$ (green points) at various times, for a continuous leak of DNAPL of Fig 5. The situation at zero time shows essentially the same behavior as the previous example with LNAPL (Fig 3). The contaminant saturation (red points) equals 0.9 for z greater than 8.0 m while it is zero in the rest of the unsaturated/saturated zone. The water saturation is different from zero only in the saturated zone, and the air saturation is 0.1 for $z > 8.0\ m$, 1 for $0\ m < z < 8.0\ m$, and zero elsewhere. After being released, the contaminant starts to migrate downward. This is represented in the middlebox of Fig 6, where the red points indicate the sharp front of DNAPL saturation moving along the vertical direction. The difference between this figure and the previous example with LNAPL (Fig 3) is that when the contaminant arrives at the groundwater table (see right-hand side of Fig 6), it keeps moving deeper into the saturated zone (that is why the red points are more immersed into the saturated zone). Notice how the sum of the three-phase saturations is always one (for a fixed depth value) and how the sharp contaminant front is immersed into the aquifer. Fig 7 shows three-dimensional numerical simulation results on the pressure as a function of the depth at different times, which is essentially similar to the previous case (in Fig 4), except that the green points here deviate from the other two (orange and blue) deeper than those in Fig 4.

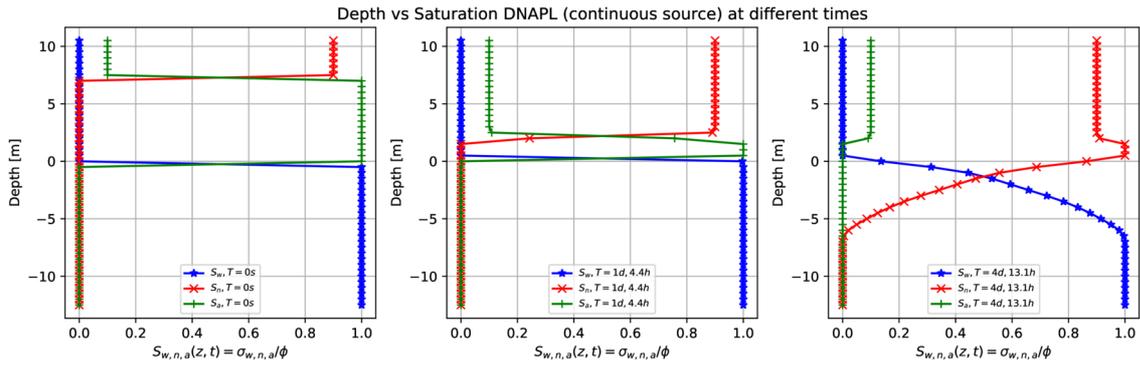

**Fig 6. Depth vs. saturation of DNAPL (continuous source) at different times.** Three-dimensional numerical simulation results of the depth as a function of the water saturation $S_w$ (blue points), DNAPL saturation $S_n$ (red points), and air saturation $S_a$ (green points) at various times for a continuous leak of DNAPL in Fig 5. Initially, at t = 0 s, a front of contaminant saturation is situated on top of the grid, which rapidly goes to zero. At the same time, it is filled by the air saturation (green point) in the unsaturated zone and water saturation in the saturated one. Notice how the sum of the three-phase saturations is always one. For later times the contaminant (red points) is immersed into the saturated zone.

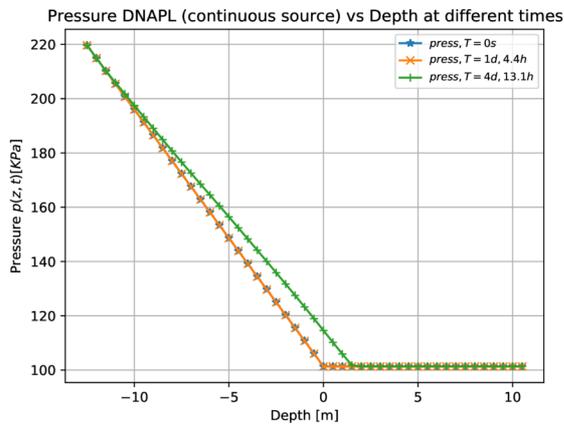

**Fig 7. Pressure vs. Depth for DNAPL (continuous source) at different times.** Three-dimensional numerical result on the pressure as a function of the depth for a continuous leak of DNAPL at different times. Initially, at t=0 s (blue points), the pressure is equal to one atmosphere in the unsaturated zone composed of air and the contaminant. Then the pressure increases as the contaminant goes downward the groundwater table to the bottom, up to a value of $220\ KPa$.

## 3.2 Migration and distribution of a small leak of NAPL

### 3.2.1 DNAPL numerical results

Consider now a small volume of leak contaminant in the unsaturated zone, a DNAPL, as is shown in Fig 8. The difference from the previous case is that the contaminant is now a finite volume (the other parameters are the same, see Table 1). Fig 8 shows three-dimensional numerical simulation results on the migration of the DNAPL for several times up to eight days and 7.1 hours. Initially, the finite volume of DNAPL of density $1200 kg/m^3$ is placed into the parallelepiped at $z = [4.0, 7.0]$ m, $x = [-5.0, 5.0]m$, and $y = [-5.0, 5.0]m$, as shown in Fig 8, first row. At later times, the contaminant moves downward through the unsaturated zone (see the second row at 5.7 hours). Due to the pressure gradient, the contaminant moves slightly to the left-hand side in the $z - x$ plane while remaining symmetric around the $y -$ axis (right-hand side). Notice how the top of the contaminant begins to empty (the DNAPL saturation contours start to change from red to blue). When it arrives at the groundwater table, it keeps going to the bottom of the saturated zone since the DNAPL is denser than the water. See the third row of Fig 8. Contextually, the top of the initial position of the contaminant is almost empty, as can be seen from its saturation contour value. The difference between this situation and that of Fig 5 is that, although it arrives at a similar depth compared with the previous case (around -10 m deep), it does not have enough inertia to move quickly in the left direction. Indeed, it takes eight days and a few hours to displace about 20 meters.

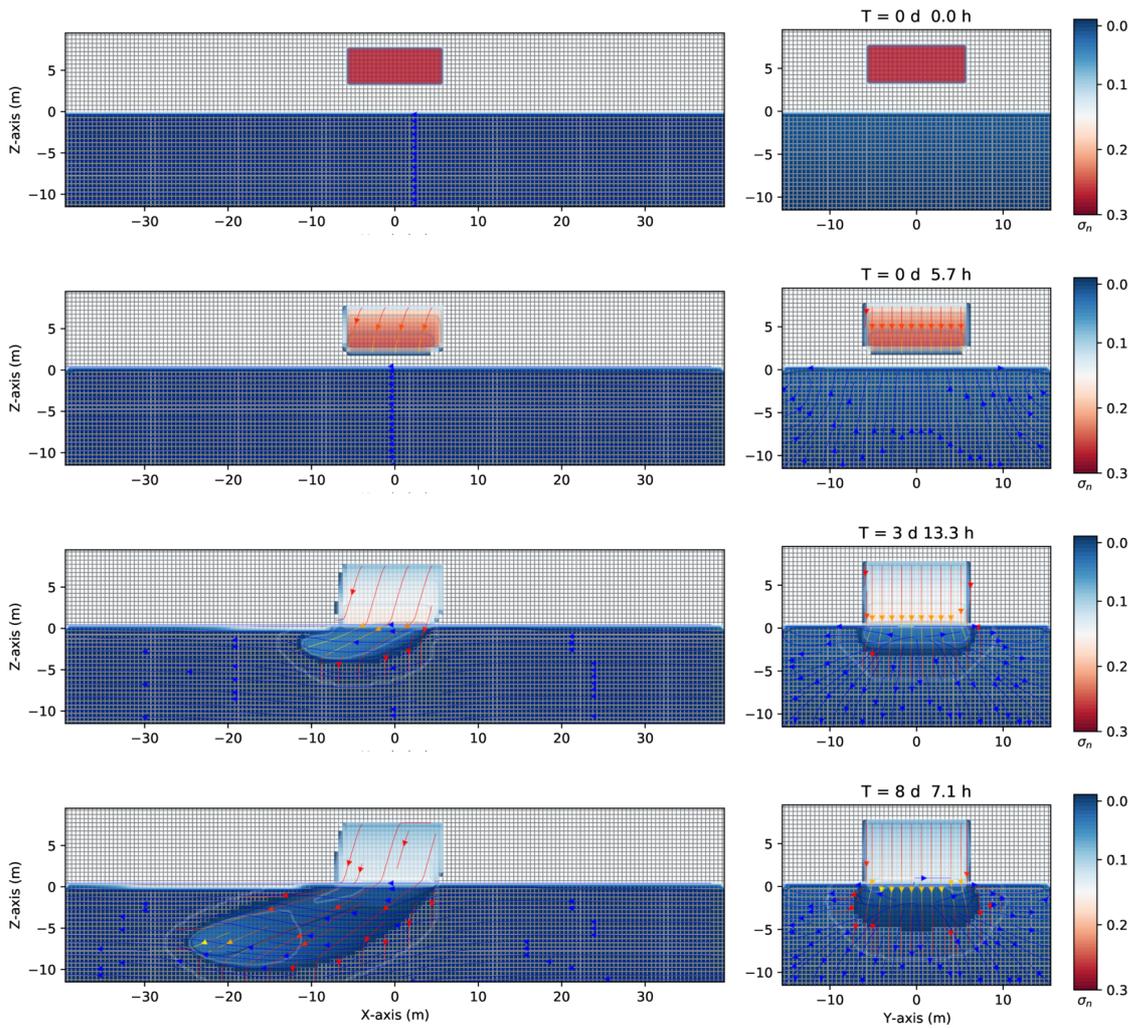

**Fig 8. Saturation contours of DNAPL (small source) at different times.** Three-dimensional numerical simulation results of the saturation contours ($\sigma_n = S_n \phi$) of a three-phase immiscible fluid flow (water + a small volume source of DNAPL + air) using a spatial grid resolution of 0.50 $m$ and a grid dimension of 80 $m$ × 32 $m$ × 22 $m$, at different times. Left-hand side shows the saturation contours in the ($z - x$) plane. Right-hand side shows the saturation contours in the ($z - y$) one. Notice how the DNAPL migrates through the saturated zone while moving in the left direction (where it is positioned a gravity at 15 degrees in the $z - x$ plane, left-hand side) at different times.

Fig 9 shows three-dimensional numerical simulation results of the depth as a function of the water saturation $S_w$ (blue points), DNAPL saturation $S_n$ (red points), and air saturation $S_a$ (green points) at various times for a small leak of DNAPL of Fig 8. At $t = 0\ s$, there is a sharp front of contaminant saturation (red points) situated in the unsaturated zone initially located at z = $[4.0, 7.0]$ m, $x = [-5.0, 5.0]m$, and $y = [-5.0, 5.0]m$. The unsaturated zone comprises the contaminant and air (green points). The saturated zone is filled up with water (blue points). Notice how the air saturation is one where there is no contaminant. Later, the contaminant moves downward, and the DNAPL saturation decreases (center). Contextually, the air phase begins to occupy the space left by the contaminant (green points). Finally, (right-hand side), the contaminant arrives at the groundwater table and enters into the saturated zone replacing the water (see the last row of Fig 8, same time). Also, it can be observed (red points) how the contaminant in the upper zone is emptied. It can be clearly seen the shock front and the rarefaction of the DNAPL saturation in the unsaturated dry zone.

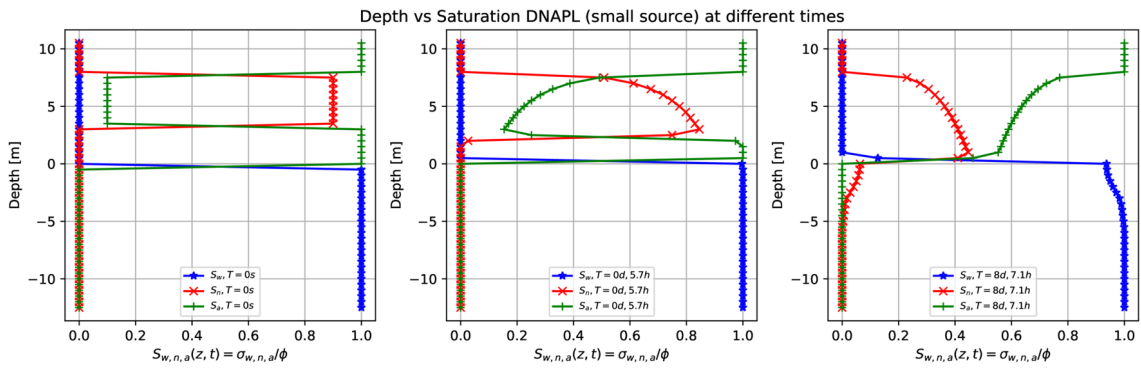

**Fig 9. Depth vs. saturation of DNAPL (small source) at different times.** Three-dimensional numerical simulation results of a depth as a function of the water saturation $S_w$ (blue points), DNAPL saturation $S_n$ (red points), and air saturation $S_a$ (green points) (plane $x = 0$) at various times for a small leak of DNAPL of Fig 8. Initially, at $t = 0\,s$, there is a sharp front of contaminant saturation (red points) situated in the unsaturated zone at z = $[4.0, 7.0]$ m. The unsaturated zone comprises the contaminant and air (green points). Later, the contaminant starts to move downward, and the saturation starts to decrease in the unsaturated zone (center). Contextually, the air phase begins to occupy the space left by the contaminant (green points). Finally, (left-hand side) the contaminant arrives at the groundwater table and enters the saturated zone (see also Fig 8, same time).

## 3.2.2 LNAPL numerical results

Consider now a small volume of contaminant, an LNAPL, with a density of $881\,kg/m^3$ (see Table 1 for details). Compared to the previous case, only the nonaqueous phase density changes. Fig 10 shows numerical simulation results of the saturation contours ($\sigma_n = S_n \phi$) of a three-phase immiscible fluid flow (water + a small volume source of LNAPL + air) using a spatial grid resolution of 0.50 $m$ and a grid dimension of 80 $m$ × 32 $m$ × 22 $m$, at different times. As before, the small leak of contaminant is released in the unsaturated zone initially located at z = $[4.0, 7.0]$ m. Fig 10 shows the situation up to nine days and 4.6 hours in the z-x plane (left-hand side) and the z-y plane (right-hand side). The small volume of LNAPL leaves the top zone entirely where it was released, being a finite amount of contaminant, and migrates downward due to the effect of the gravity force (see second the third row). When it arrives at the groundwater table (third row), it remains in it, and a few parts of this contaminant interacts with the capillary fringe. Notice a difference to Fig 8, where the inertia from a constant contaminant source extends in the capillary fringe zone. In this case, a small volume of contaminant LNAPL floats on the groundwater table while slowly moving to the left-hand side. After nine days, has traveled for about 25 meters in the left direction (fourth row, left-hand side). Also, notice how the flow lines in the saturated zone remain almost undisturbed even by the presence of the LNAPL. Compared with Fig 8 (small volume of DNAPL), the LNAPL arrives later at the groundwater table and empties first with respect to the previous case.

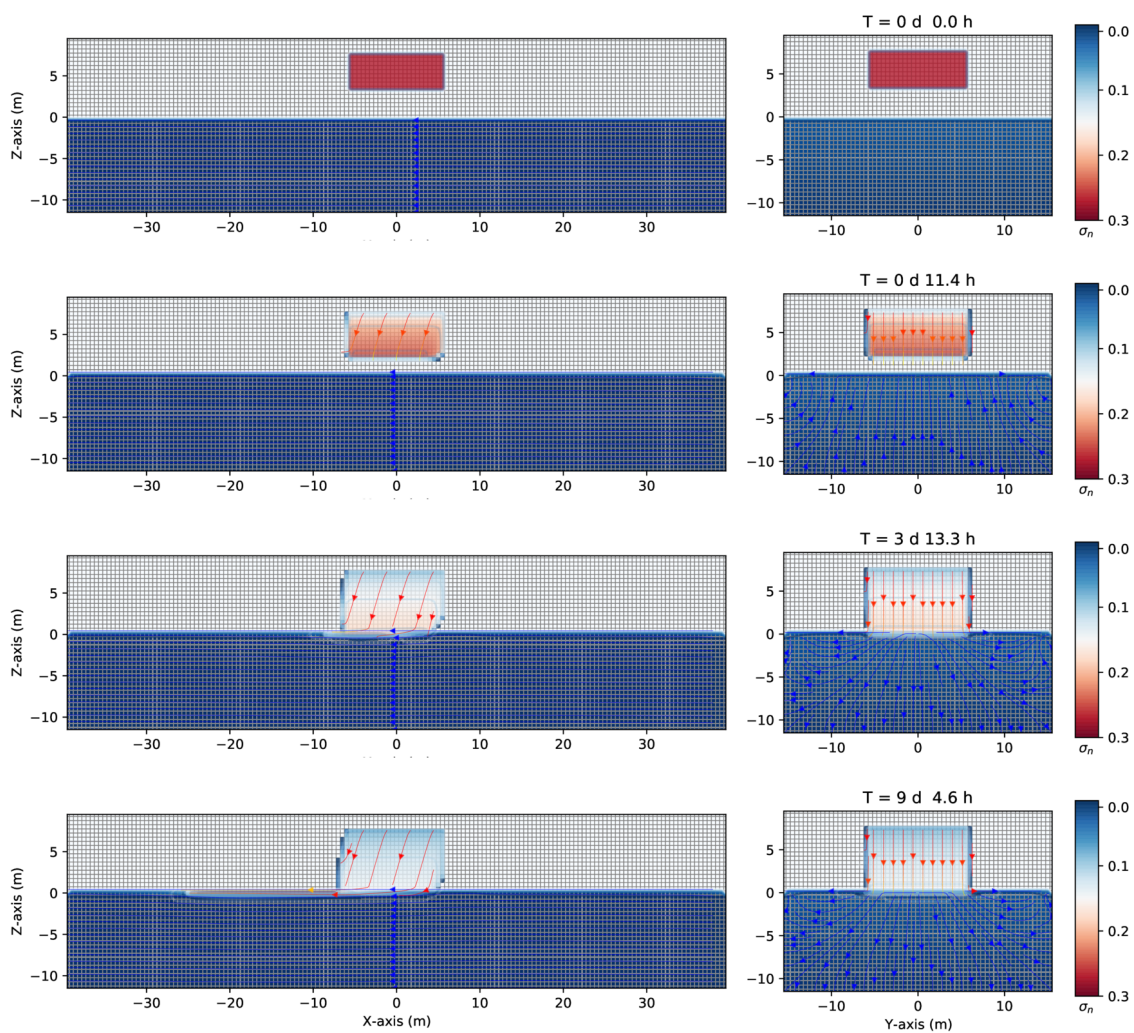

**Fig 10. Saturation contours of LNAPL (small source) at different times.** Three-dimensional numerical simulation results of the saturation contours ($\sigma_n = S_n \phi$) of a three-phase immiscible fluid flow (water + a small volume source of LNAPL + air) using a spatial grid resolution of 0.50 $m$ and a grid dimension of 80 $m$ × 32 $m$ × 22 $m$, at different times. Notice how the LNAPL migrates through the saturated zone while moving in the left direction due to a pressure gradient and remains entirely on the capillary fringe zone.

The same situation is corroborated in Fig 11, which shows numerical simulation results of the depth as a function of the water saturation $S_w$ (blue points), LNAPL saturation $S_n$ (red points), and air saturation $S_a$ (green points) at various times for a small leak of LNAPL of Fig 10. Substantially, the principal difference between Fig 9 and Fig 11 is that for the DNAPL, the contaminant (red points) goes through the saturated zone (bellow the groundwater table) for the LNAPL does not happen; therefore the saturation empties less.

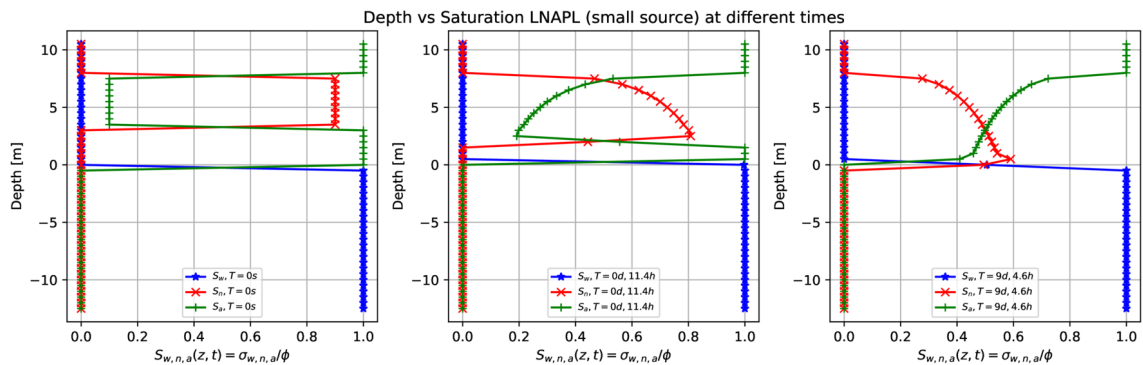

**Fig 11. Depth vs. saturation of LNAPL (small source) at different times.** Numerical simulation results of a depth as a function of the water saturation $S_w$ (blue points), LNAPL saturation $S_n$ (red points), and air saturation $S_a$ (green points) at various times for a small leak of LNAPL of Fig 10. Initially, at t = 0 s, there is a sharp front of contaminant saturation (red points) situated in the unsaturated zone at z = [4.0,7.0] m. Later, the contaminant moves downward, and the saturation decreases (center). On the right-hand side, the contaminant reaches the groundwater table and does not enter the saturated zone except in a small quantity.

# 3.3 Migration and distribution of a continuous leak of DNAPL in presence of impermeable "lenses" in heterogeneous aquifers

## 3.3.1 Impermeable "lens" in the unsaturated zone

Consider now the case in Fig 12 where a continuous leak source of DNAPL (similar to the case represented in Fig 5) encounters an impermeable zone (such as a clay lens in heterogeneous aquifer media), here simplified as a parallelepiped with an absolute permeability of $4.14 \times 10^{-14} m^2$ (ten thousand smaller than the value of the rest of the domain, see Table 1). The (green) impermeable parallelepiped is situated in the unsaturated zone, at $z = [2.0, 6.0]m, x = [-10.0, +10.0]m, y = [-10.0, +10.0]m$. The grid dimension and spatial grid are similar to the previous cases. Once the continuous source of DNAPL, initially situated at $z = 8.0$ m, $x = [-5.0, 5.0]m$, and $y = [-5.0, 5.0]m$, is released, it goes downward and encounters the impermeable zone after 5.7 hours. See Fig 12 first row, where the three-dimensional numerical simulation results of the saturation contours ($\sigma_n = S_n\phi$) of a three-phase immiscible fluid flow (water + a continuous source of DNAPL + air) are shown. Afterward, the DNAPL (left-hand side) moves around the impermeable zone and preferably flows to the left side under the gravity's action positioned 15 degrees left to the z-direction (second and third row, left-hand side). While flows symmetrically in the y-direction (right-hand side). Since the DNAPL is much denser than the water, it goes downward through the saturated zone (similar to Fig 5). Part of the mobile DNAPL arrives at the other side of the parallelepiped, situated at $x = 10\ m$. See fourth-row. The right-hand side of Fig 12 shows a symmetric behavior where the contaminant moves around the impermeable zone and finally arrives at the saturated one. Instead, the left-hand side, shows the contaminant that reaches $x = -38\ m$ and $10\ m$ depth after seven days and 2.7 hours.

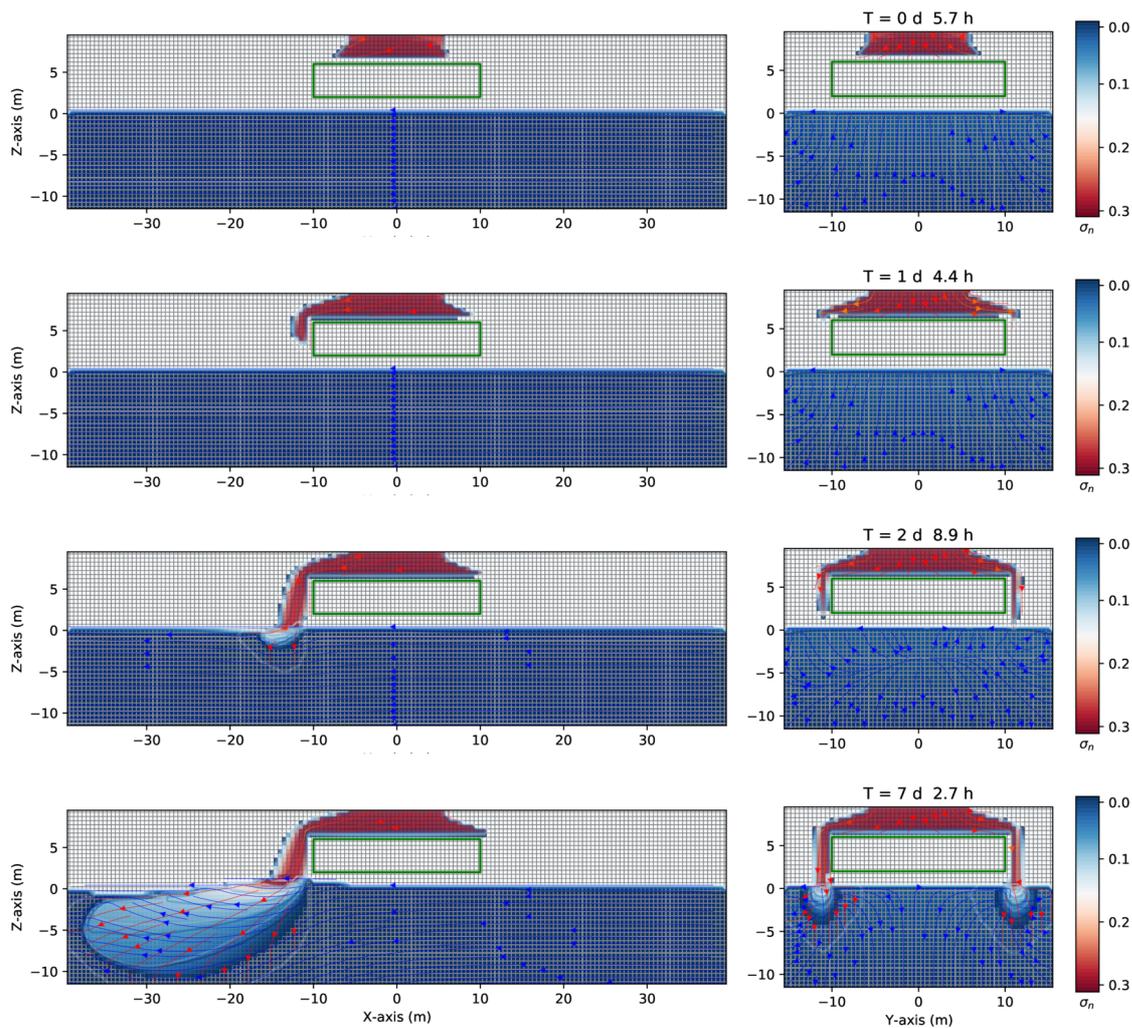

**Fig 12. Saturation contours of DNAPL (with an impermeable lens) at different times.** Three-dimensional numerical simulation results of the saturation contours ($\sigma_n = S_n \phi$) of a three-phase immiscible fluid flow (water + a continuous source of DNAPL + air) using a spatial grid resolution of $0.50\ m$ and a grid dimension of $80\ m \times 32\ m \times 22\ m$, at different times. The DNAPL leak encounters an impermeable obstacle (depicted in green) situated at $z = [2.0, 6.0]\ m, x = [-10.0, +10.0]\ m, y = [-10.0, +10.0]\ m$. After seven days and 2.7 hours the contaminant has reached the saturated (aquifer) zone at $x = -10\ m$ depth.

Fig 13 shows three-dimensional numerical simulation results of a depth as a function of the water saturation $S_w$ (blue points), DNAPL saturation $S_n$ (red points), and air saturation $S_a$ (green points) at various times for a continuous leak of DNAPL of Fig 12 (at plane $x = 0$). The continuous source of DNAPL leak encounters an impermeable obstacle (depicted in green in Fig 12). Initially, a sharp front of contaminant saturation (red points) is situated in the unsaturated zone at z = [8.0,10.0] m. See the left-hand side (similar to the previous case in Fig 5). After two days and 8.9 hours, the contaminant moves downward, and its saturation (red points) increases from 0.9 up to one, just above the impermeable at $z = 6.0\ m$, as it accumulates in that area (center figure). Then it saturation abruptly goes to zero since we are showing the plane $x = 0$, and instead, the contaminant moves on the left-hand side (following the groundwater flow). This situation remains almost invariable for later times (right-hand side).

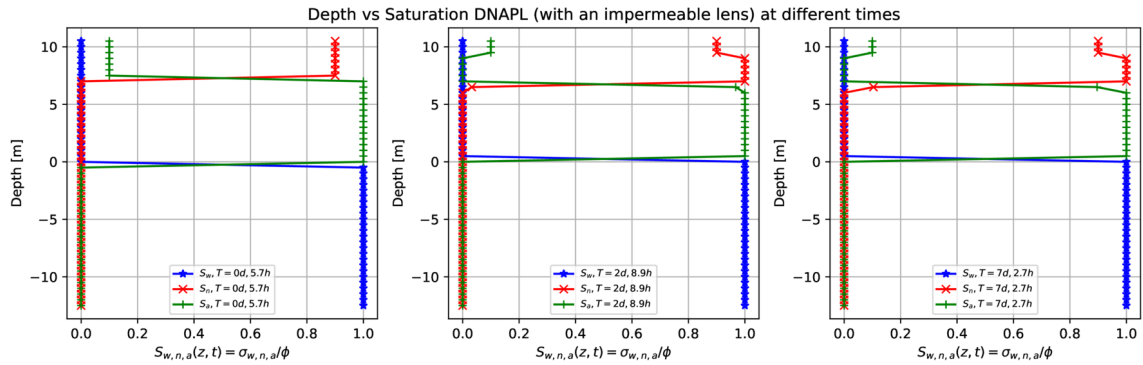

**Fig 13. Depth vs. saturation of DNAPL (with an impermeable lens) at different times.** Three-dimensional numerical simulation results of a depth as a function of the water saturation $S_w$ (blue points), DNAPL saturation $S_n$ (red points), and air saturation $S_a$ (green points) at various times for a continuous leak of DNAPL of Fig 12 (plane $x = 0$). The DNAPL leak encounters an impermeable obstacle (depicted in green in Fig 12). The left-hand side shows a sharp front of contaminant saturation (red points) situated in the unsaturated zone at z = $[8.0,10.0]$ m. At later times, (center) after two days and 8.9 hours, the contaminant saturation increases to 1.0 due to an accumulation on top of the impermeable zone while part of it reaches the saturated zone. This situation remains almost invariable for later times (right-hand side) since we plot the $x = 0$ plane.

Fig 14 shows a similar case to that of Fig. 12, where the impermeable obstacle (depicted in green) is a smaller one, and is situated at $z = [2.0, 4.0]m, x = [-5.0, +5.0]m, y = [-5.0, +5.0]m$. The three-dimensional numerical simulation results of the contour saturation in the planes $z - x$ and $z - y$ are shown at different times. Since the impermeable parallelepiped is a smaller one, with respect to the previous case, the continuous source of DNAPL has the opportunity to reach the groundwater table zone faster than the previous case and also from the other side of the parallelepiped situated at $x = 5\ m$ (see Fig 14, second row). Once arrived at the groundwater table/capillary fringe, the DNAPL keeps going downward and moving to the left direction due to a pressure gradient, as in the previous cases (see the third row). The last row of Fig 14 shows the saturation contour of the contaminant after seven days and 2.7 hours. Both concentrations situated below the parallelepiped are moving together toward the left side.

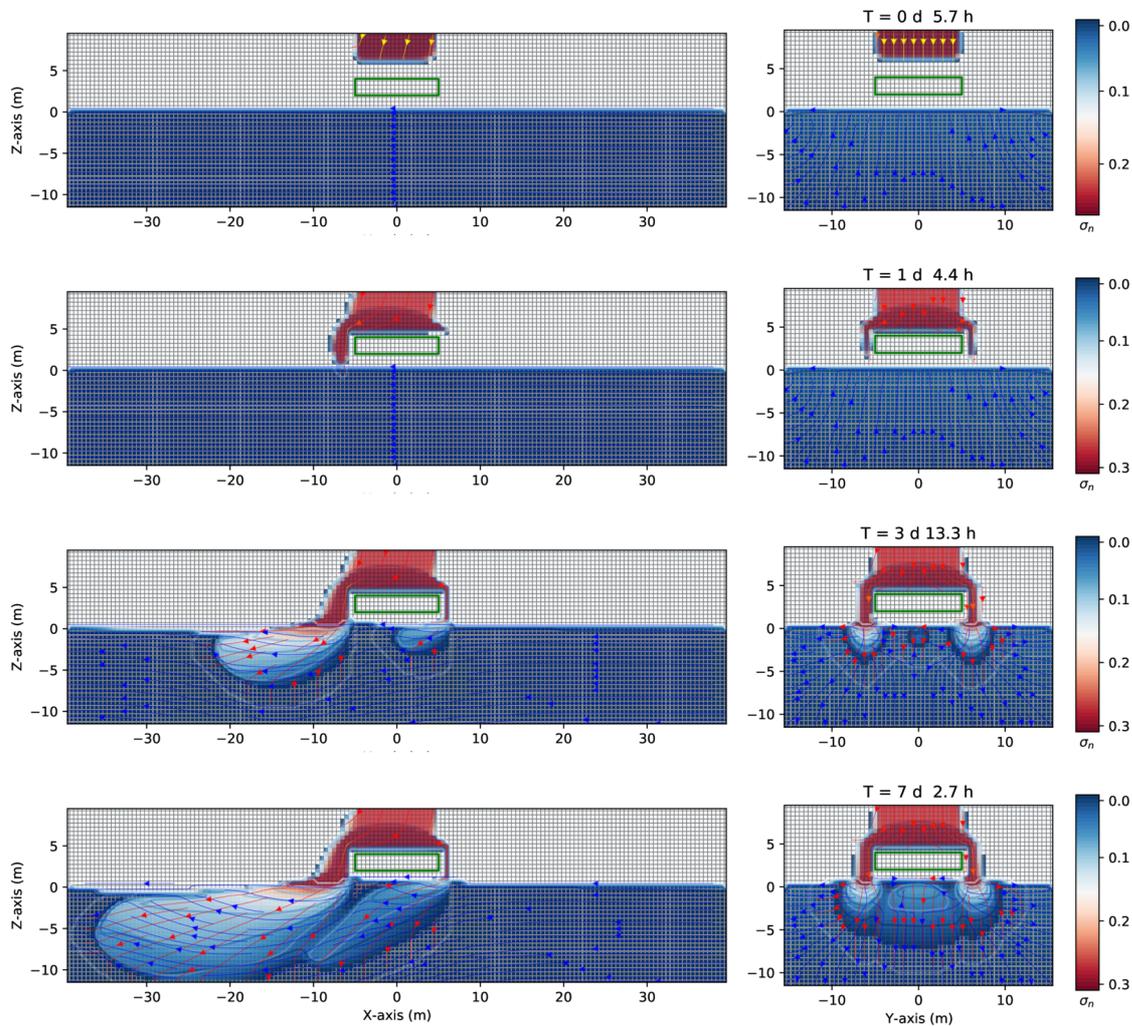

**Fig 14. Saturation contours of DNAPL (with a smaller impermeable lens) at different times.** Three-dimensional numerical simulation results of the saturation contours ($\sigma_n = S_n\phi$) of a three-phase immiscible fluid flow (water + a continuous source of DNAPL + air) using a spatial grid resolution of $0.50\ m$ and a grid dimension of $80\ m\ \times\ 32\ m\ \times\ 22\ m$, at different times (plane $x = 0$). The continuous source of DNAPL encounters an impermeable obstacle (depicted in green), smaller than the one in Fig 12, situated at $z = [2.0,4.0]m, x = [-5.0, +5.0]m, y = [-5.0, +5.0]m$. After arriving at the groundwater table, the continuous source of DNAPL keeps going downward and moving to the left direction on both sides.

Fig 15 shows the three-dimensional numerical simulation results of a depth as a function of the water saturation $S_w$ (blue points), DNAPL saturation $S_n$ (red points), and air saturation $S_a$ (green points) at various times for a continuous leak of DNAPL of Fig 14 (plane $x = 0$). The continuous source of DNAPL encounters an impermeable lens (depicted in green), as shown in Fig 14, smaller than the one in Fig 12. After 5.7 hours (left-hand side) there is a sharp front of contaminant saturation (red points) situated in the unsaturated zone at $z = [8.0, 10.0]$ m. The rest of the unsaturated zone is filled with air (green points), while the saturated zone is filled with water (blue points). At later times, three days and 13.3 hours, the contaminant has already arrived at the impermeable lens and accumulates on top of it (middle). That is why the saturation sharp front increases from 0.9 to 1.0 (red points) and then goes abruptly to zero (in the region where there is the impermeable). On the right-hand side, there is the DNAPL saturation (red points) different from zero just below the groundwater table and is the DNAPL that passes under the impermeable in the plane x=0. See Fig 14 fourth-row.

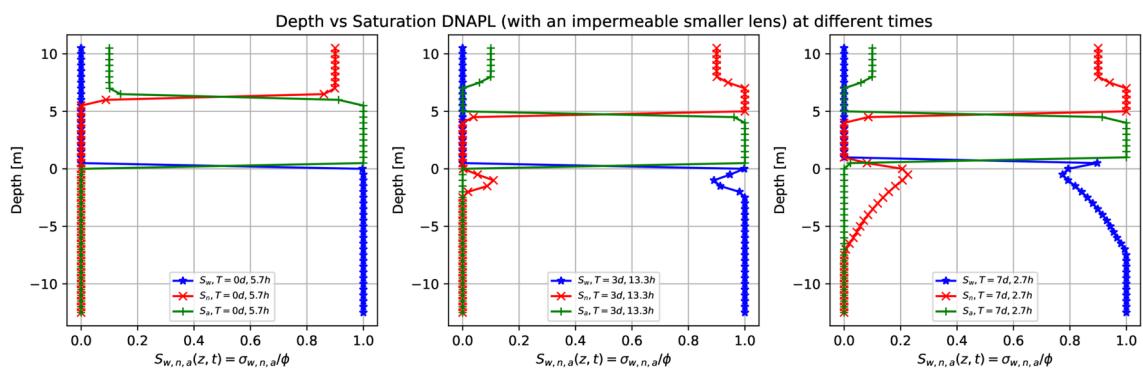

**Fig 15. Depth vs. saturation of DNAPL (with a smaller impermeable lens) at different times.** Three-dimensional numerical simulation results of a depth as a function of the water saturation $S_w$ (blue points), DNAPL saturation $S_n$ (red points), and air saturation $S_a$ (green points) at various times for a continuous leak of DNAPL of Fig 14 (plane x=0). The DNAPL leak encounters an impermeable obstacle (depicted in green in Fig 14). After 5.7 hours, there is a sharp front of contaminant saturation (red points) situated in the unsaturated zone at $z = [8.0, 10.0]$ $m$. The DNAPL saturation increases from 0.9 to one at later times since it accumulates on top of the impermeable parallelepiped. Then abruptly goes to zero and finally increases in the saturated zone (right-hand side). Notice that the DNAPL and water saturation sum is one for a fixed depth value.

Fig 16 shows the three-dimensional visualization of the numerical simulation results of the saturation contaminant iso-surface of Fig 14, after 8.15 days (just a little further than the time in the fourth column of Fig 14). The figure was generated using an open-source post-processing named VisIt (https://wci.llnl.gov/simulation/computer-codes/visit). Notice how the contaminant moves on the impermeable parallelepiped zone and spills around the impermeable zone. The figure shows the iso-surface of equal contaminant density distribution after 8.15 days of the evolution shown in Fig 14.

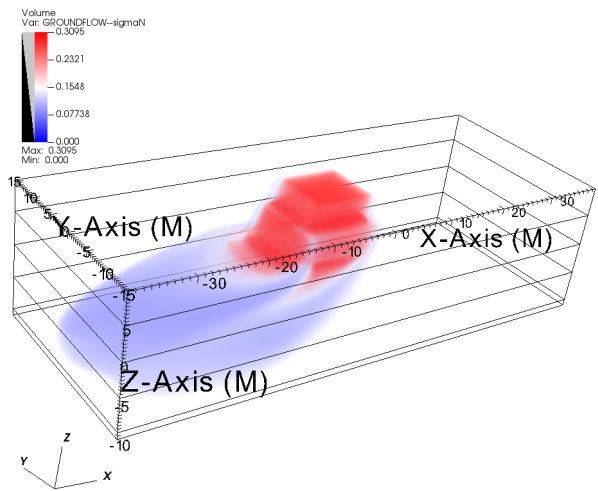

**Fig 16. Saturation iso-surface of DNAPL (with an impermeable lens) at time 8.15 days.** The three-dimensional visualization of the numerical simulation results of the contaminant saturation iso-surface at time 8.15 days. Notice the impermeable parallelepiped zone just below the contaminant released on top of the grid. This figure was generated using VisIt (an open-source post-processing) and the three-dimensional saturation data of Fig 14.

## 3.3.2 Impermeable "lens" in the saturated zone

Suppose that the impermeable obstacle is situated in the saturated zone (instead of the unsaturated one). See Fig 17 where a continuous source of DNAPL is released in the unsaturated zone, similar to Fig 12, but the impermeable "lens" (depicted in green) is located at the coordinates, $z = [-5.0, -2.0]m, x = [-10.0, +10.0]m, y = [-10.0, +10.0]m$. Once the continuous source of DNAPL arrives at the saturated zone encounters an impermeable obstacle, therefore it turns around and continues its direction toward the left-hand side. Fig 17 shows the numerical simulation results of the saturation contours ($\sigma_n = S_n\phi$) of a three-phase immiscible fluid flow (water + a continuous source of DNAPL + air) using a spatial grid resolution of $0.50\ m$ and a grid dimension of $80\ m \times 32\ m \times 22\ m$, at different times. The DNAPL encounters an impermeable obstacle (depicted in green) in the saturated zone (see the first row). After arriving at the groundwater table (see the second row), the DNAPL keeps going downward and moving to the left direction, due to a pressure gradient (see the third row), and eventually will reach the bottom zone being denser than the water (fourth row).

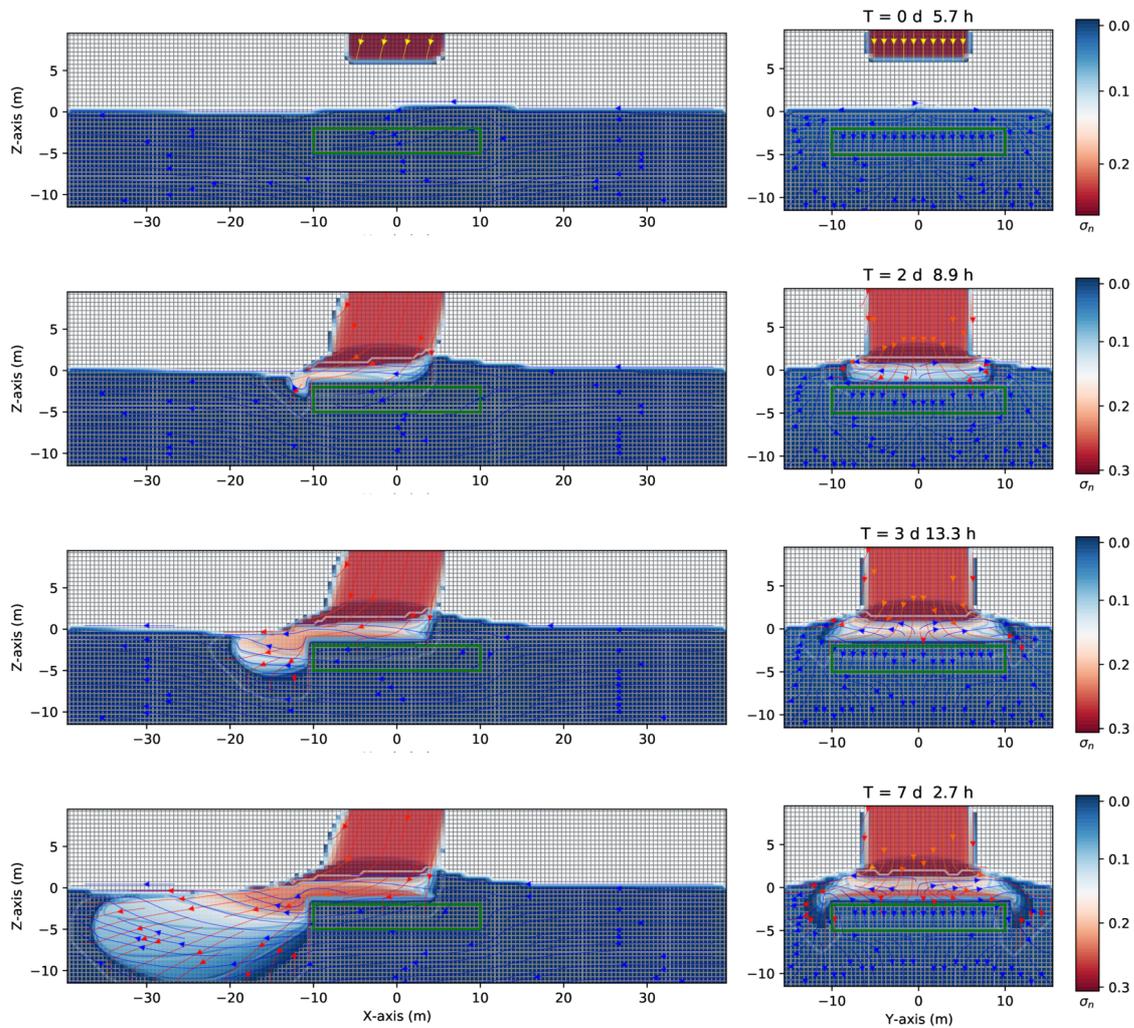

**Fig 17. Saturation contours of DNAPL (with an impermeable in the groundwater) at different times.** Three-dimensional numerical simulation results of the saturation contours ($\sigma_n = S_n \phi$) of a three-phase immiscible fluid flow (water + a continuous source of DNAPL + air) using a spatial grid resolution of $0.50\ m$ and a grid dimension of $80\ m\ \times\ 32\ m\ \times\ 22\ m$, at different times. The DNAPL encounters an impermeable obstacle (depicted in green), situated at $z = [-5.0, -2.0]\ m$, $x = [-10.0, +10.0]\ m$, $y = [-10.0, +10.0]\ m$.

Fig 18 shows the numerical simulation results of a depth as a function of the water saturation $S_w$ (blue points), DNAPL saturation $S_n$ (red points), and air saturation $S_a$ (green points) at various times for a continuous leak source of DNAPL of Fig 17 (plane $x = 0$). The DNAPL encounters an impermeable lens (depicted in green) in the saturated zone, as shown in Fig 17. The left-hand side shows the situation after 5.7 hours where the continuous source of DNAPL is situated on top of the parallelepiped, similar to Fig 15. After two days and 8.9 hours (middle), the contaminant arrives at the groundwater table (see Fig 17, second row). Its saturation increases from 0.9 to 1.0 since it arrives at the saturated zone and tends to accumulate in that region, but at the same time keeps moving to the bottom (see red points), being denser than the water. Since the impermeable lens is situated just below the groundwater table, the contaminant saturation first decreases and then increases again. It finally goes to zero. This situation is better represented on the right-hand side after seven days and 2.7 hours.

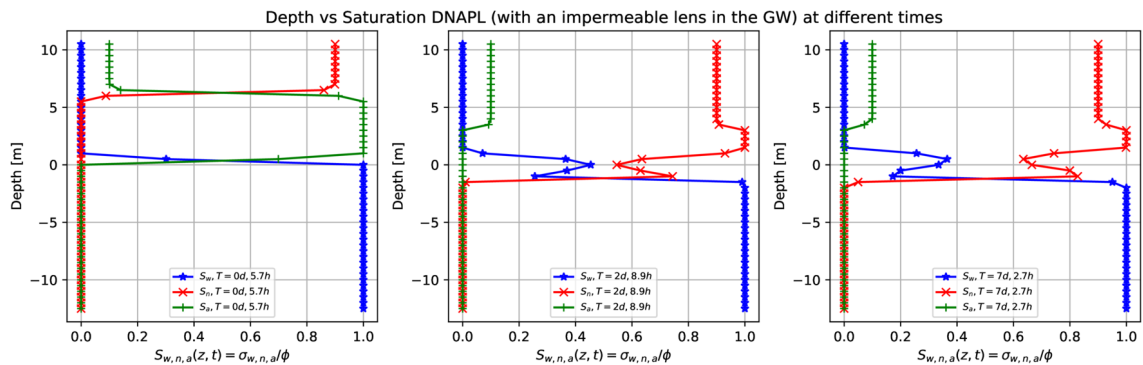

**Fig 18. Depth vs. saturation of DNAPL (with an impermeable lens in the groundwater) at different times** Three-dimensional numerical simulation results of a depth as a function of the water saturation $S_w$ (blue points), DNAPL saturation $S_n$ (red points), and air saturation $S_a$ (green points) at various times for a continuous leak of DNAPL of Fig 17 (plane $x = 0$). The continuous source of DNAPL encounters an impermeable obstacle (depicted in green in Fig 17). Initially, the contaminant (red points) is situated on the parallelepiped (left-hand side). Then the sharp front starts to move downward (middle) until it arrives at the saturated zone where the impermeable obstacle causes the contaminant to accumulate just below the groundwater table (right-hand side).

# 4. Conclusions

In this work, we presented high-resolution three-dimensional numerical modeling investigation of the migration of three-phase immiscible fluid flow in variably saturated zones. We investigate the temporal evolution of the migration of the immiscible contaminant problem in a porous medium following the saturation contour profiles of the three-phases fluids flow. We considered both light nonaqueous phase liquid and dense nonaqueous phase liquid, initially released in unsaturated dry soil, and investigated several initial conditions, including impermeable parallelepipeds that mimic clay "lenses" in heterogeneous aquifer systems. The numerical simulations were obtained using CactusHydro code [18], based on the Cactus toolkit [36-37], that uses a high-resolution shock-capturing conservative method that precisely follows the advective part of the fluid flow. The results were validated through a classical convergence test running the same code at different resolutions.

We show that it is possible to follow with high precision the migration of a contaminant sharp front (LNAPL or DNAPL, continuous or small volume source) in a variably saturated zone as a whole using a unique mathematical model which includes the complete set of mathematical equations expressed in terms of the saturation, permeability, capillary pressure, density, the viscosity of the three-phases. We show that the difference between the fate of a DNAPL and LNAPL (when the other parameters are the same) is just the density of the contaminant. We also show the numerical results of the three-phase saturation very precisely as a function of the depth. Next step of this research will be to compare numerical simulations results with laboratory experimental results such as a contaminant leakage in a sand tank.